\documentclass[bibyear]{aa} 

\usepackage{graphicx}
\usepackage{txfonts}
\usepackage{natbib}
\bibpunct{}{}{,}{a}{}{,} 
\begin{document}
%
\def\hi {H\,{\sc i}}
\def\hii {H\,{\sc ii}}
\def\water {H$_2$O}
\def\meth {CH$_{3}$OH}
\def\dg{$^{\circ}$}
\def\kms{km\,s$^{-1}$}
\def\ms{m\,s$^{-1}$}
\def\jyb{Jy\,beam$^{-1}$}
\def\mjyb{mJy\,beam$^{-1}$}
\def\solmass {\hbox{M$_{\odot}$}}
\def\solum {\hbox{L$_{\odot}$}} 
\def\d {$^{\circ}$}
\def\n {$n_{\rm{H_{2}}}$}
\def\kmsg{km\,s$^{-1}$\,G$^{-1}$}
\def\tbo {$T_{\rm{b}}\Delta\Omega$}
\def\tb {$T_{\rm{b}}$}
\def\om{$\Delta\Omega$}
\def\dvi {$\Delta V_{\rm{i}}$}
\def\dvz {$\Delta V_{\rm{Z}}$}
\def\code {FRTM code}
\title{Rapidly increasing collimation and magnetic field changes of a protostellar \water ~maser outflow}

\author{G.\ Surcis  \inst{1}
  \and 
  W.H.T. \ Vlemmings \inst{2}
 \and
  H.J. \ van Langevelde \inst{1,3}
 \and
  C. \ Goddi \inst{1}
  \and
  J.M. \ Torrelles  \inst{4}
  \and
  J. \ Cant\'{o} \inst{5}
  \and
  S. \ Curiel \inst{5}
  \and
  S.-W. \ Kim \inst{6}
  \and
  J.-S. \ Kim \inst{7}
    }

\institute{
 Joint Institute for VLBI in Europe, Postbus 2, 79990 AA Dwingeloo, The Netherlands
 \and
 Chalmers University of Technology, Onsala Space Observatory, SE-439 92 Onsala, Sweden
 \and
 Sterrewacht Leiden, Leiden University, Postbus 9513, 2300 RA Leiden, The Netherlands
 \and
 Institut de Ci\`encies de l'Espai (CSIC)-Institut de Ci\`encies del Cosmos (UB)/IEEC, E-08028 Barcelona, Spain
 \and
 Instituto de Astronom\'{i}a (UNAM), Apdo Postal 70-264, Cd. Universitaria, 04510-M\'{e}xico D.F., M\'{e}xico 
 \and
 Korea Astronomy and Space Science Institute, 776 Daedeokdaero, Yuseong, Daejeon 305-348, Republic of Korea
 \and
 National Astronomical Observatory of Japan, 2-21-1 Osawa, Mitaka, Tokyo 181-8588, Japan}
\date{Received ; accepted}
\abstract
{W75N(B) is a massive star-forming region that contains three radio continuum sources (VLA\,1, VLA\,2, and VLA\,3), which are thought to
be three massive young stellar objects at three different evolutionary stages. VLA\,1 is the most evolved and VLA\,2 the least evolved source. 
The 22~GHz \water ~masers associated with VLA\,1 and VLA\,2 have been mapped at several epochs over eight years. While the \water ~masers in VLA\,1 
show a persistent linear distribution along a radio jet, those in VLA\,2 are distributed around an expanding shell. Furthermore, \water ~maser
polarimetric measurements revealed magnetic fields aligned with the two structures.}
{Using new polarimetric observations of \water ~masers, we aim to confirm the elliptical
expansion of the shell-like structure around VLA\,2 and, at the same time, to determine if the magnetic fields around the two 
sources have changed.}
{The NRAO Very Long Baseline Array was used to measure the linear polarization and the Zeeman-splitting of the 22~GHz \water ~masers
towards the massive star-forming region W75N(B).}
{The \water ~maser distribution around VLA\,1 is unchanged from that previously observed. We made an elliptical fit of the \water ~masers around VLA\,2. 
We find that the shell-like structure is still expanding along the direction parallel to the thermal radio jet of VLA\,1. While the magnetic field around 
VLA\,1 has not changed in the past $\sim$7 years, the magnetic field around VLA\,2 has changed its orientation according to the new direction of the 
major-axis of the shell-like structure and it is now aligned with the magnetic field in VLA\,1. }
{}


\maketitle
\section{Introduction}
The formation of massive stars and the evolution of associated protostellar outflow is still a matter of debate (e.g., Beuther \&
Shepherd \cite{beu05}; Zinnecker \& Yorke \cite{zin07}). Beuther \& Shepherd (\cite{beu05}) propose an evolutionary scenario in which 
well-collimated outflows occur 
in the very early phases of high-mass star formation (HMSF) and, in their evolution, the outflows get progressively less collimated because 
of the build-up of an \hii ~region. Recent magnetohydrodynamics (MHD) simulations show that magnetic fields coupled to  
prestellar disks drive outflows, which could also be poorly collimated at very early stages of HMSF, depending on the magnetic field strength
(e.g., Banerjee \& Pudritz \cite{ban07}, Seifried et al. \cite{sei11, sei12}). \\
\indent Although multi-epoch Very Long Baseline Interferometry (VLBI) observations of 22~GHz \water ~masers were successful in identifying 
jets/outflows (Goddi et al. \cite{god05}; Moscadelli et al. \cite{mos07}; Sanna et al. \cite{san10}), monitoring studies of  
outflow formation and magnetic field evolution at early stages of HMSF are still lacking. Fortunately, one very singular case where we can do both
studies at the same time does exist; this is W75N(B).\\
\indent The active massive star-forming region W75N(B) is located at a distance of $1.3$~kpc (Rygl et al. 
\cite{ryg12}) that contains three massive young stellar objects (YSOs) within an area of $\sim1.5'' \times 1.5''$ ($\sim 2000~ \rm{au} \times 2000~ \rm{au}$), 
named VLA\,1, VLA\,2, and VLA\,3 (Torrelles et al. \cite{tor97}; Carrasco-Gonz\'alez et al. \cite{gon10}). The sources VLA\,1 and VLA\,3 show elongated 
radio continuum emission consistent with a
thermal radio jet, while VLA\,2, which is located between VLA\,1 and VLA\,3, shows unresolved continuum emission ($\leq 0.08''$) of unknown nature 
(Torrelles et al. \cite{tor97}). The three sources are thought to be YSOs at three different evolutionary stages; in particular, VLA\,1 is the 
most evolved and VLA\,2 the least evolved (Torrelles et al. \cite{tor97}). Several maser species have been detected towards W75N(B) (e.g., Baart et al. 
\cite{baa86}; Torrelles et al. \cite{tor97}; Surcis et al. \cite{sur09}). In particular, the 22~GHz \water ~masers have been monitored over a period 
of eight years from 1999 to 2007 (e.g., Torrelles et al. \cite{tor03}, hereafter T03; Surcis et al. \cite{sur11}, hereafter S11; Kim et al. \cite{kim13}, 
hereafter K13). \\
\indent Remarkably, while the \water ~masers in VLA\,1 trace a collimated thermal radio jet of $\sim1''$ (1300~au) with  
$\rm{PA_{jet}}\approx+43$\d ~(Torrelles et al. \cite{tor97}), those around VLA\,2 are tracing an expanding shell that evolved 
from a quasi-spherical to a collimated structure over eight years (T03, S11, K13). 
Moreover, S11 analyzed the polarized emission of 22~GHz \water ~masers and found that the magnetic field around VLA\,1 and VLA\,2 (separated
by just 1300~au) has different orientation and strength.\\
\indent Therefore, we propose W75N(B) as the best case known where the transition from a non-collimated to a well-collimated outflow in the very 
early phase of HMSF can be observed in ``real time''. In this letter, we present new polarimetric VLBI observations of \water ~masers to confirm the elliptical
expansion of the shell-like structure around VLA\,2 as well as to determine possible changes in the magnetic field.
\begin{figure*}[ht!]
\centering
\includegraphics[width = 8 cm]{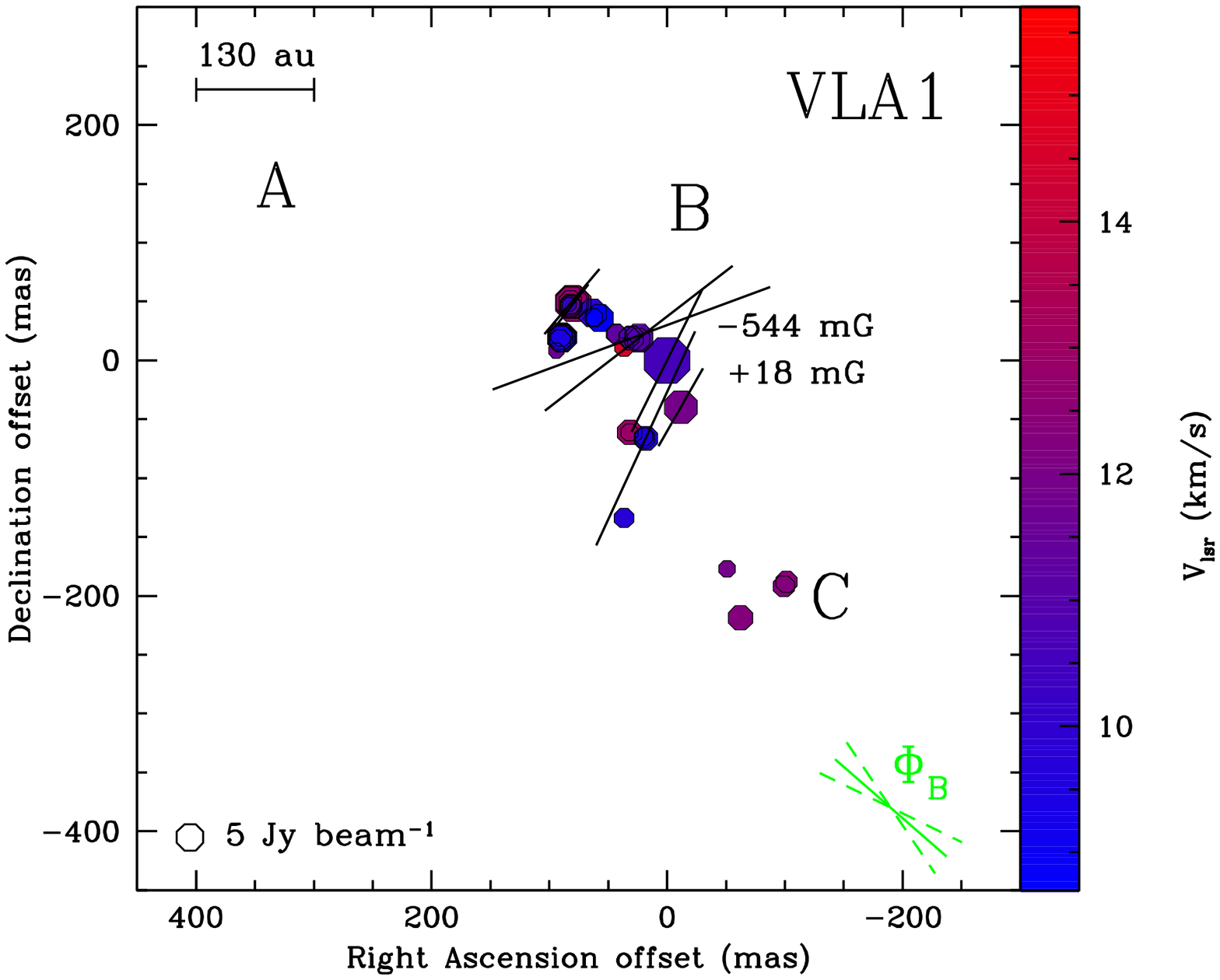}
\includegraphics[width = 8 cm]{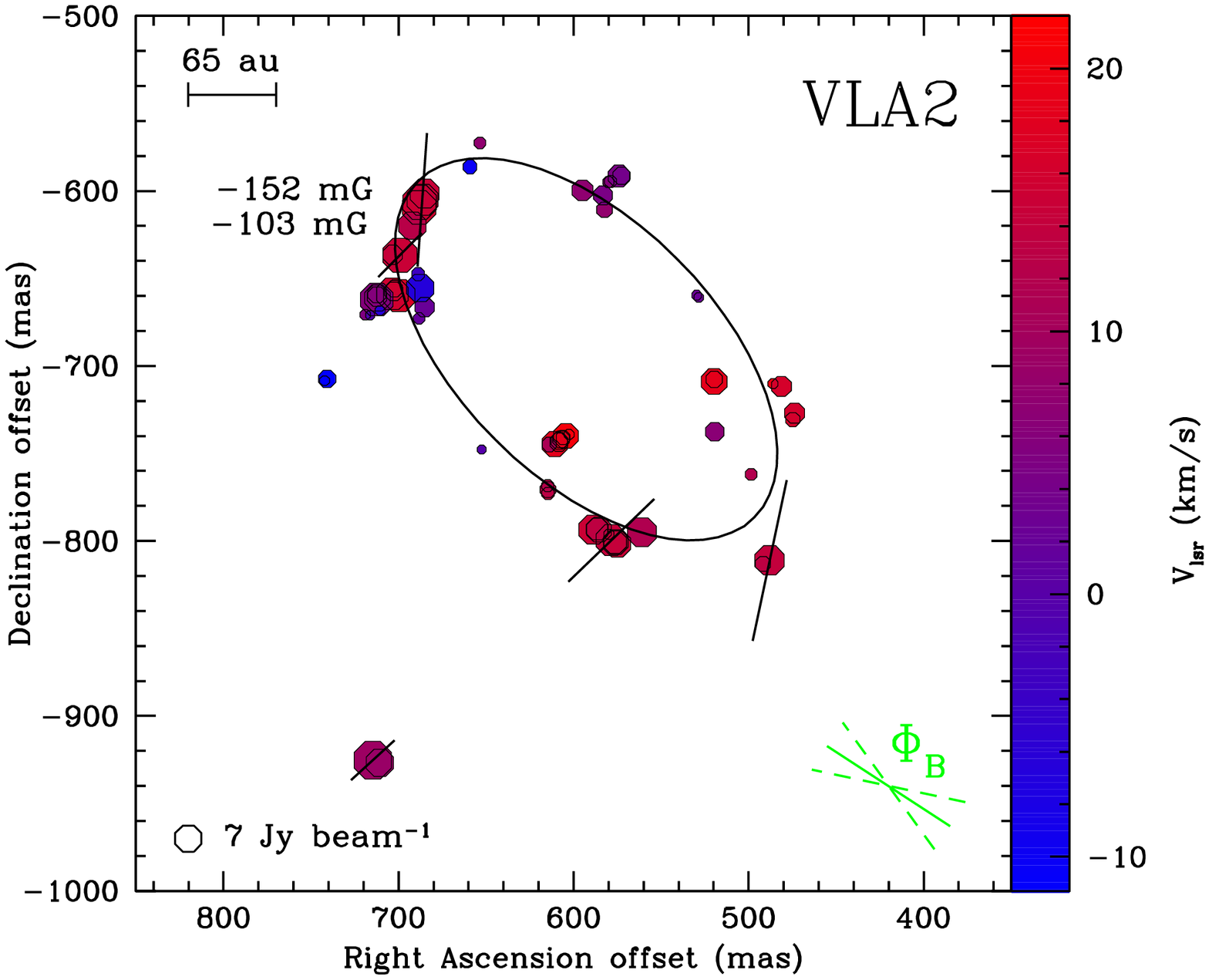}
\caption{A close-up view of the 22~GHz \water ~maser features detected around the radio source VLA\,1 (left panel) and VLA\,2 (right panel).
The reference position is $\alpha_{2000}=20^{\rm{h}}38^{\rm{m}}36^{\rm{s}}\!.435$ and 
$\delta_{2000}=42^{\circ}37'34''\!\!.84$ (see Sect.~\ref{obsana}). The octagonal symbols 
are the identified maser features in the present work scaled logarithmically according to their peak flux density (Tables~\ref{VLA1_wat}
and \ref{VLA2_wat}, \textit{online material}). The linear polarization vectors, scaled logarithmically according to polarization fraction 
$P_{\rm{l}}$, are overplotted. In the bottom-right
corner of both panels the error-weighted orientation of the magnetic field ($\Phi_{\rm{B}}$, see Sect.~\ref{res}) is also reported; 
the two dashed segments indicate the uncertainties. The ellipse drawn in the right panel is the result of the 
best fit of the \water ~masers detected in the present work (epoch 2012.54). Its parameters are listed in Table~\ref{ellipse}. The estimated
values of the magnetic field strength are also shown in both panels next to the corresponding \water ~maser.}
\label{water}
\end{figure*}
\section{Observations and analysis}
\label{obsana}
The star-forming region W75N(B) was observed in the $6_{16}-5_{23}$ transition of \water ~(rest frequency: 22.23508~GHz) with the 
NRAO\footnote{The National Radio Astronomy Observatory (NRAO) is a facility of the National Science Foundation operated under 
cooperative agreement by Associated Universities, Inc.} VLBA on July 15, 2012. The observations were made in full polarization
mode using a bandwidth of 4~MHz to cover a velocity range of $\sim54$~\kms. The data were correlated with the DiFX correlator 
using 2000 channels and generating all four polarization combinations (RR, LL, RL, LR) with a spectral resolution of
2~kHz ($\sim$0.03~\kms). Including the overheads, the total observation time was 8~hr.\\
\indent The data were calibrated using AIPS by following the same calibration procedure described in S11. We used the same calibrator 
used by S11, i.e., J2202+4216. Then we imaged the \textit{I}, \textit{Q}, \textit{U}, and \textit{V} cubes ($\rm{rms}=21$~\mjyb) 
using the AIPS task IMAGR (beam size 0.87~mas~$\times$~0.61~mas,  $\rm{PA}=+3.75$\d). The \textit{Q} and \textit{U} cubes were combined to produce 
cubes of polarized intensity (\textit{POLI}) and polarization angle ($\chi$). Because W75N(B) was observed 11 days after a POLCAL observations run made by 
NRAO\footnote{http://www.aoc.nrao.edu/$\sim$smyers/calibration/}, we calibrated the linear polarization 
angles of the \water ~masers by comparing the linear polarization angle of J2202+4216 that we measured with the angles measured 
during that POLCAL observations run ($\chi_{\rm{J2202+4216}}=-15$\d$\!\!.0\pm0$\d$\!\!.3$). The formal errors on $\chi$ are 
due to thermal noise. This error is given by $\sigma_{\chi}=0.5 ~\sigma_{P}/P \times 180^{\circ}/\pi$ (Wardle \& Kronberg 
\cite{war74}), where $P$ and $\sigma_{P}$ are the polarization intensity and corresponding rms error, respectively. We estimated the 
absolute position of the 
brightest maser feature through fringe rate mapping by using the AIPS task FRMAP. As the formal errors of FRMAP are $\Delta\alpha=2.6$~mas 
and $\Delta\delta=1.2$~mas, the absolute position uncertainty will be dominated by the phase fluctuations. We estimate these to be on the order of no
more than a few mas from our experience with other experiments and varying the task parameters.\\
\indent We analyzed the polarimetric data following the procedure reported in S11. First, we identified the \water ~maser features and determined
the linear polarization fraction ($P_{\rm{l}}$) and $\chi$ for each identified \water ~maser feature. Second, we used the full radiative transfer 
method (FRTM) code for 22~GHz \water ~masers (Vlemmings et al. \cite{vle06}; Appendix~\ref{appA}, \textit{online material}). 
The output of this code provides 
estimates of the emerging brightness temperature (\tbo) and of the intrinsic thermal linewidth (\dvi).
From \tbo ~and $P_{\rm{l}}$, we then determined the angle between the maser propagation direction and the magnetic field ($\theta$).
If $\theta>\theta_{\rm{crit}}=55$\d ~the magnetic field 
appears to be perpendicular to the linear polarization vectors; otherwise, it is parallel (Goldreich et al. \cite{gol73}).
Finally, the best estimates of \tbo ~and \dvi ~are included in the FRTM code to produce the $I$ and $V$ models used 
for measuring the Zeeman splitting (see Appendix~\ref{appA}). 
\section{Results}
\label{res}
\subsection{VLA\,1}
The \water ~masers in VLA\,1 are distributed along the radio jet as previously observed by T03 (epoch~1999.25), S11 (epoch~2005.89), and 
K13 (epoch~2007.41). Surcis et al. (\cite{sur11}) found the \water ~masers clustered in three groups, named A, B, and C. In this work 
(epoch 2012.54), we detected 38 
\water ~masers (named VLA1.01 -- VLA1.38; Table~\ref{VLA1_wat}, \textit{online material}) in groups~B and C, but not in A (Fig.~\ref{water}). 
Group~A was also not detected in 1999 and 2007. 
For a detailed comparison of the \water ~maser parameters measured in 
epochs~2005.89 and 2012.54 see Table~\ref{wat0512} (\textit{online material}).\\
\indent We detected linearly polarized emission from seven \water ~masers ($P_{\rm{l}}=0.6\%-4.5\%$), and the error-weighted linear
polarization angle is $\langle\chi\rangle_{\rm{VLA1}}=-41^{\circ}\pm15^{\circ}$. The FRTM code was able to fit four out 
of the seven \water ~masers (Table~\ref{VLA1_wat}). Because the lower limit of the fitting range of \tbo ~is $10^6~\rm{K~sr}$, 
the estimated values of \dvi ~and \tbo ~are upper limits. The error-weighted values of the outputs are 
$\langle\Delta V_{\rm{i}}\rangle_{\rm{VLA1}}<2.4$~\kms, $\langle T_{\rm{b}}\Delta\Omega\rangle_{\rm{VLA1}}<10^{6}$~K~sr, and 
$\langle \theta \rangle_{\rm{VLA1}}=+90$\d$^{+10^{\circ}}_{-10^{\circ}}$. This implies that the magnetic field is perpendicular 
to the linear polarization vectors and the error-weighted orientation on the plane of the sky is 
$\langle \Phi_{\rm{B}}\rangle_{\rm{VLA1}}=+49^{\circ}\pm15$\d. The foreground, ambient, 
and internal Faraday rotations are small or negligible as shown by S11.  \\
\indent Circularly polarized emission is detected in VLA1.06 ($P_{\rm{V}}=0.07\%$) and VLA1.12 ($1.8\%$). Because the FRTM code
was not able to determine \tbo ~and \dvi ~for VLA1.12, we considered the values of the closest maser VLA1.10 to produce the 
$I$ and $V$ models (Fig.~\ref{circpol}, \textit{online material}). The estimated magnetic field strengths along the line of sight 
($B_{||}$) are +18~mG and -544~mG (a negative magnetic field strength indicates that the magnetic field is pointing towards 
the observer; otherwise away from the observer). The magnetic field strength $B$ is related to $B_{||}$ by $B_{||}=B~\rm{cos} \theta$
if $\theta\neq\pm90$\d. Because $\langle \theta \rangle_{\rm{VLA1}}={90^{\circ}}^{+10^{\circ}}_{-10^{\circ}}$, we can only provide a
lower limit of $B$ for VLA\,1 (Table~\ref{wat0512}).
\begin{figure*}[ht!]
\centering
\includegraphics[width = 8 cm]{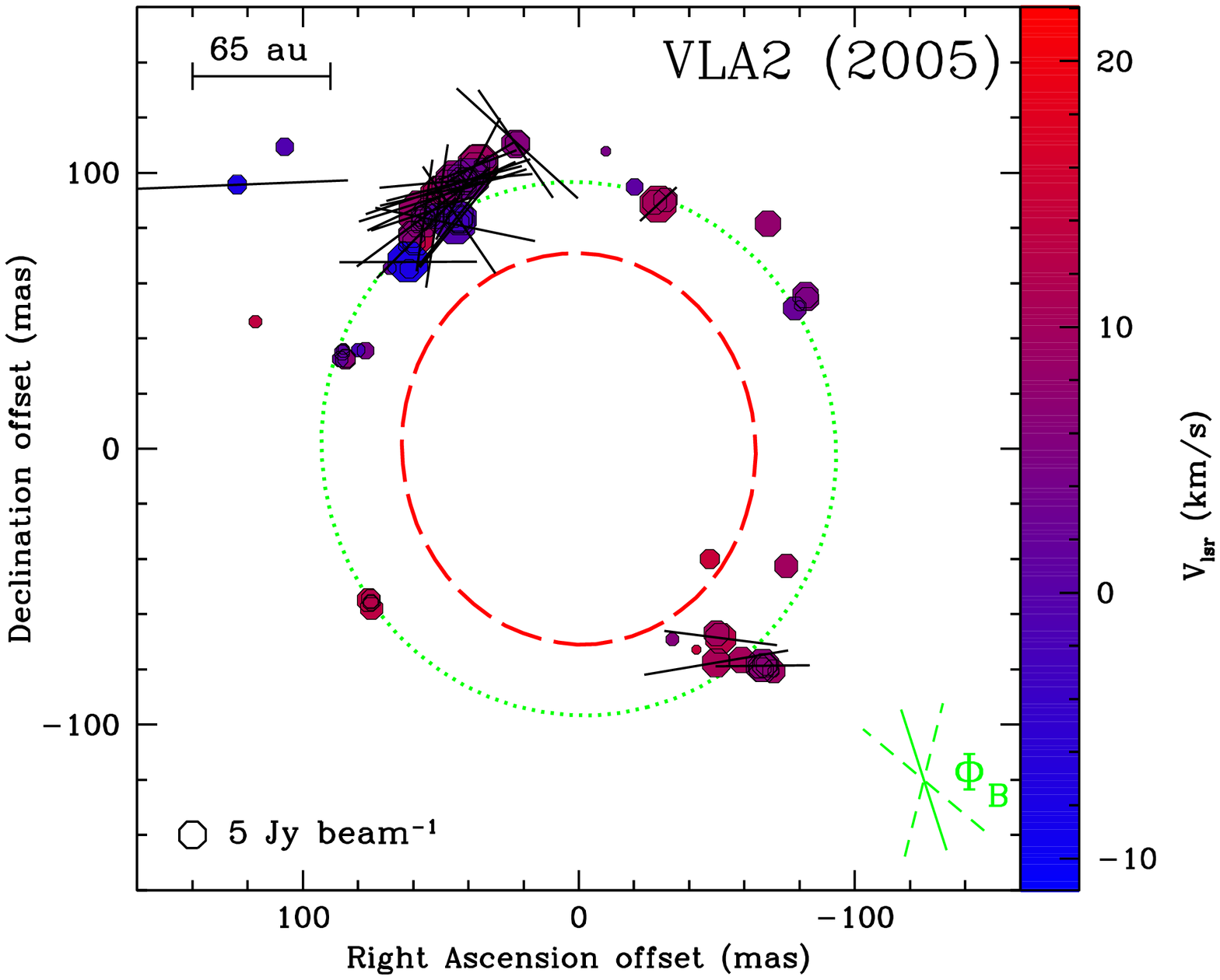}
\includegraphics[width = 8 cm]{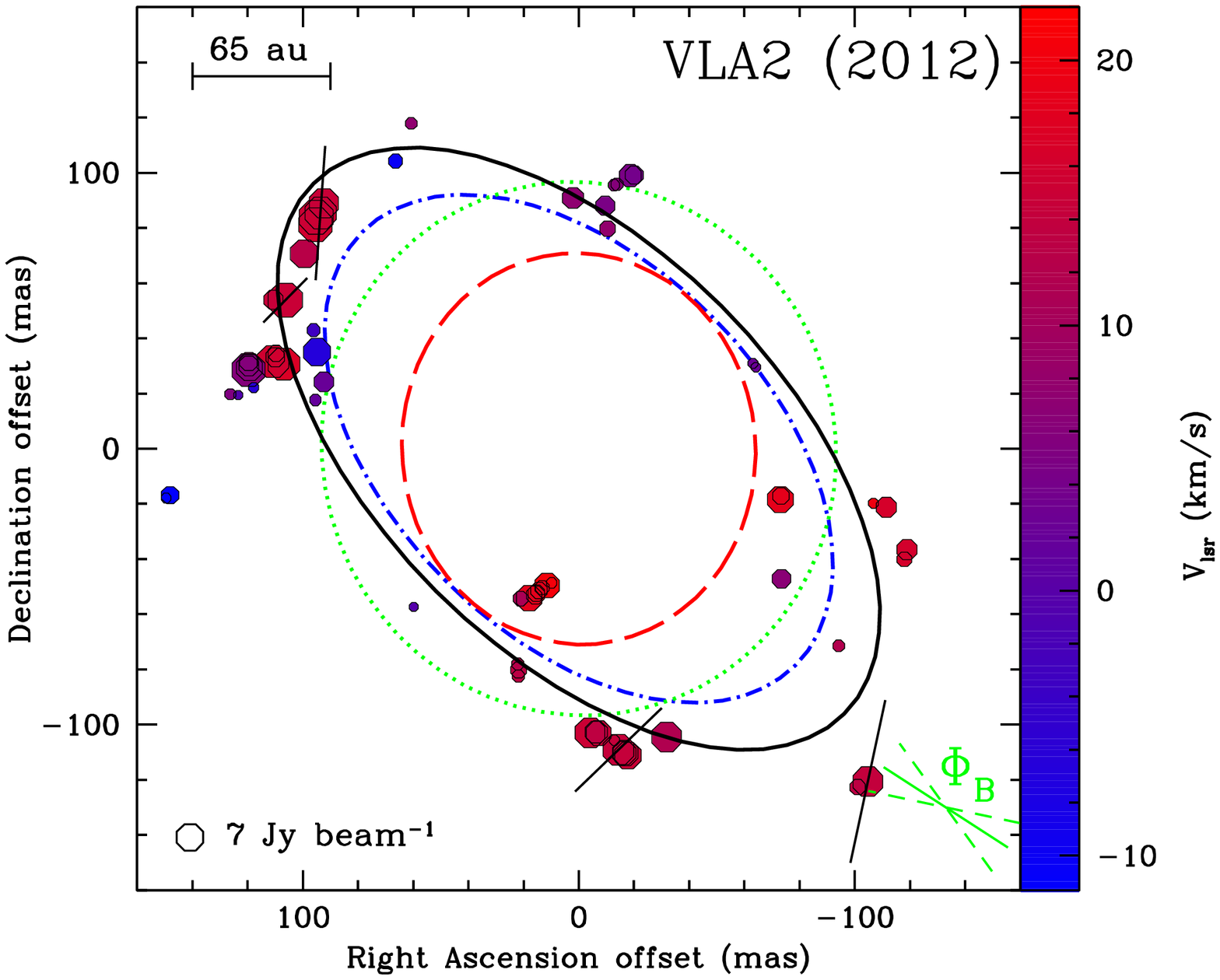}
\caption{Comparison of the \water ~masers around VLA\,2 in epoch 2005.89 (left panel; S11) and in epoch 2012.54 (right panel; 
present work). A comparison of the elliptical fits of the \water ~maser distributions observed in the past 13 years is also
shown (see Fig.~\ref{water} for more details). The maser LSR radial velocity 
bar on the right of both panels shows the same velocity range. Four ellipses are drawn, which are assumed to have the same center 
(the (0,0) reference position). They are the results of
the best fit of the \water ~masers detected by T03 (epoch 1999.25; red dashed ellipse), S11 (epoch 2005.89; green dotted ellipse), 
K13 (epoch 2007.41; blue dot-dashed ellipse), and the present work (epoch 2012.54; black solid ellipse). Their parameters are listed
in Table~\ref{ellipse}.
}
\label{vla2}
\end{figure*}
\subsection{VLA\,2}
We detected 68 \water ~masers (named VLA2.01--VLA2.68; Table~\ref{VLA2_wat}, \textit{online material}) showing an elliptical 
distribution similar to that observed in epoch~2007.41 (K13). An elliptical
fit reveals that the semi-major axis ($a$) and the semi-minor axis ($b$) are $136\pm4$~mas and $73\pm2$~mas, respectively, and 
the position angle is $\rm{PA}=45$\d$\pm2$\d. The center of the ellipse is at the position $\rm{c_{\alpha}} = +593\pm2$~mas, 
$\rm{c_{\delta}}= -690\pm3$~mas with respect to VLA1.06. The eccentricity, $e=\sqrt{1-(b/a)^2}$, of the fitted ellipse is $0.84\pm0.05$.\\
\indent Five \water ~masers show linearly polarized emission ($P_{\rm{l}}=0.7\%-1.6\%$), and the error-weighted linear polarization
angle is $\langle\chi\rangle_{\rm{VLA2}}=-33^{\circ}\pm21^{\circ}$. The FRTM code was able to properly fit only VLA2.64 and the outputs
are \dvi$_{,\rm{VLA2}}=1.98$~\kms, \tbo$_{\rm{VLA2}}=6 \times 10^{8}$~K~sr, and 
$\theta_{\rm{VLA2}}=+84$\d$^{+6^{\circ}}_{-10^{\circ}}$. This implies that the magnetic field is perpendicular 
to the linear polarization vectors and the error-weighted orientation on the plane of the sky is 
$\langle \Phi_{\rm{B}}\rangle_{\rm{VLA2}}=+57^{\circ}\pm21$\d. \\
\indent Circularly polarized emission was detected towards two \water ~masers, namely VLA2.44 ($P_{\rm{V}}=0.7\%$) and VLA2.48 
($P_{\rm{V}}=0.4\%$). These masers do not show linear polarization and consequently no information on \dvi ~and \tbo ~is available. In
order to measure the magnetic field strength, we decided to assign values to \dvi ~and \tbo ~that could produce the best $I$ and 
$V$ fitting models. These are \dvi$~=2.0$~\kms ~for both masers, and \tbo$~=5\times10^{9}$~K~sr and \tbo$~=10^{9}$~K~sr for VLA2.44 
and VLA2.48, respectively. The goodness of the fit can be seen in Fig.~\ref{circpol}. The estimated $B_{||}$ are -152~mG and -103~mG.
\section{Discussion}
\subsection{The immutable VLA\,1}
The \water ~masers in VLA\,1 show a linear distribution ($\rm{PA}\approx43$\d) persistent over 13 years. Nevertheless, there are minor 
differences compared to S11. Specifically, the flux density has generally decreased from 2005 to 2012 (Table~\ref{wat0512}). This may
explain the disappearance of the masers of group~A, which also had larger $V_{\rm{lsr}}$ than groups B and C and thus they
were probably tracing an occasional fast ejection event ($V_{\rm{lsr}}^{\rm{VLA\,1}}=9$~\kms, Carrasco-Gonz\'alez et al. \cite{gon10}). 
The inferred magnetic field in VLA\,1 is along the radio jet and it is almost 
aligned with the large-scale CO-outflow ($\rm{PA_{out}}=66$\d; Hunter et al. \cite{hun94}), as measured in 2005 (Table~\ref{wat0512}).\\
 \indent The stability of the maser and magnetic field distribution around VLA\,1 might indicate a relatively evolved stage of 
 this massive YSO in comparison with VLA\,2 (see below).
\begin {table}[th!]
\caption []{Comparison of the fitted parameters of the ellipses from K13 (1999.25, 2005.89, 2007.41) and the present work (2012.54).} 
\begin{center}
\scriptsize
\begin{tabular}{ l c c c c c c }
\hline
\hline
\,\,\,\,\,(1)&(2)    & (3)       & (4)      & (5)                      & \multicolumn{2}{c}{(6)}          \\
 Epoch   & $a$       & $b$      & PA        & $e$\tablefootmark{a}     & \multicolumn{2}{c}{Expansion Velocity\tablefootmark{b}}  \\
         & (mas)     & (mas)    & (\d)      &                          &  ($\rm{mas~yr^{-1}}$) & (\kms)   \\ 
\hline
 1999.25 & $71\pm1$\tablefootmark{c}  & $64\pm1$\tablefootmark{c} & $5\pm3$\tablefootmark{c}   & $0.43\pm0.01$            &                       &         \\
 2005.89 & $97\pm3$  & $93\pm2$ & $15\pm45$ & $0.28\pm0.02$            & $3.9\pm0.5$           & $24\pm3$ \\
 2007.41\tablefootmark{c} & $111\pm1$& $68\pm1$  & $45\pm1$ & $0.79\pm0.02$ & $9.2\pm2.0$           & $57\pm12$\\
 2012.54 & $136\pm4$ & $73\pm2$ & $45\pm2$  & $0.84\pm0.05$            & $4.9\pm0.8$           & $30\pm5$ \\
        &           &          &           &                          & $4.9\pm0.3$\tablefootmark{d} & $30\pm2$\tablefootmark{d} \\
\hline
\end{tabular} 
\end{center}
\tablefoot{
\tablefoottext{a}{Eccentricity, $e=\sqrt{1-(b/a)^2}$.}
\tablefoottext{b}{From the difference in the semi-major 
axis size of the ellipse between different epochs (1999.25--2005.89; 2005.89--2007.41; 2007.41--2012.54). }
\tablefoottext{c}{The considered epoch is May 29, 2007.}
\tablefoottext{d}{Between epoch 1999.25 and epoch 2012.54.}
}
\label{ellipse}
\end{table}
\subsection{The evolution of the expanding \water ~maser shell in VLA\,2}
Unlikely VLA\,1, VLA\,2 has shown remarkable evolution both in structure and magnetic field in 
the last decade, as probed by the \water ~masers mapped with VLBI at four different epochs. 
In all epochs, the \water ~masers have shown a different distribution around 
VLA\,2, in size and/or shape, going from circular (T03, S11) to elliptical (K13, present work; Fig.~\ref{vla2} and 
Table~\ref{ellipse}).\\
\indent In epoch~1999.25, the elliptical 
fit reveals that $a$ and $b$ have almost the same value ($e=0.43\pm0.01$, Table~\ref{ellipse}) indicating that the \water ~masers 
are tracing an almost circular shell-like structure (T03). This shell is thought to be the signature of a shock caused by the expansion of 
a non-collimated outflow; T03 also measured the proper motion of the individual \water ~masers, concluding that they are moving 
outward from VLA\,2 at $\sim19$~\kms.\\
\indent In epoch~2005.89, S11 found that the circular shell increased its size by about 30~mas, but it did not changed its 
shape significantly ($e=0.28\pm0.02$). 
In about six years the circular shell expanded with a velocity of $24\pm3$~\kms ~that is consistent with the proper 
motions of the individual \water ~masers (T03). This suggests that the formation of an early non-collimated outflow from a 
massive YSO is observed at mas scale; S11 also determined that the magnetic field is of the order of $1-2$~G around VLA\,2 and it is 
oriented along $a$.\\
\indent After only two years from the observations of S11, K13 observed that the \water ~maser shell is still expanding, but along a more
dominant axis with $\rm{PA}=+45$\d$\pm1$\d ~($e=0.79\pm0.02$). The increment of the ellipticity could be the sign of the launching of a 
collimated jet that overtakes the non-collimated outflow. Surprisingly, the shell is now aligned with both the thermal radio jet and the
magnetic field in VLA\,1. \\
\indent Our observations of epoch 2012.54 show that the expansion of the shell still continues after five years and that its ellipticity 
has increased ($e=0.84\pm0.02$). The position angle of our fit is equal to that determined by K13 indicating that the supposed launching
of a collimated jet has actually happened (Table~\ref{ellipse}).\\
\indent In contrast to the magnetic field in VLA\,1, the magnetic field in VLA\,2 has changed its orientation substantially 
(Fig.~\ref{vla2}). The magnetic field has rotated by about +40\d ~during the past seven years and it is now aligned with the major axis of 
the fitted ellipse of epoch~2012.54 ($\rm{PA}=+45$\d$\pm2$\d). By comparing $\langle \Phi_{\rm{B}}\rangle_{\rm{VLA2}}$ with  
$\langle \Phi_{\rm{B}}\rangle_{\rm{VLA1}}$, we notice that the magnetic fields around VLA\,2 and VLA\,1 are now aligned with both the jet
in VLA\,1 and the elliptical \water ~maser shell in VLA\,2. This configuration may arise if the large-scale magnetic field of
W75N(B) drives the orientation of the two jets and potentially regulates HMSF as suggested by recent observations 
(Girart et al. \cite{gir09}, Tan et al. \cite{tan13}).
A test of this hypothesis may be to determine the morphology of the 
magnetic field of the region at large scale via dust polarization observations. 
Incidentally, we note that the inferred magnetic field direction also appears to be perpendicular to the filamentary core and 
its velocity gradient traced by NH$_{3}$ thermal emission (Carrasco-Gonz\'alez et al. \cite{gon10}).\\
\indent A possible physical framework to explain our results in VLA\,2 may be provided by recent MHD simulations (Seifried et al. \cite{sei12}).
In this context, the magnetic pressure drives a slow non-collimated outflow 
in the very first phase of protostellar formation. Immediately after the formation of a Keplerian disk, a short-lived fast and collimated jet overtakes 
the slow outflow. This could be qualitatively in agreement with our findings in VLA\,2.\\
\indent In addition, a comparison between
$\overline{|B_{||}|}_{\rm{VLA2}}^{2005.89}$=345\,\,mG and $\overline{|B_{||}|}_{\rm{VLA2}}^{2012.54}$=128\,\,mG shows 
that the magnetic field in epoch~2012.54 is one third of the magnetic field measured in epoch~2005.89. The masers at the two epochs probe 
different gas properties and the measured variation of the magnetic field could simply be a consequence of it. 
We thus speculate that the variation may be due to the launching of the fast jet, but present simulations do not include the variation of 
the magnetic field strength during the early outflow evolution to corroborate our hypothesis.\\
\indent From an observational perspective, to confirm our scenario it is necessary to monitor the expanding motion of the 22~GHz \water ~maser 
structure and the magnetic field evolution in the region over time. Furthermore, the determination of the 3D velocity 
structure of the outflow obtained with new proper motion measurements of the \water ~masers and of the evolution of the continuum 
morphology of VLA\,2 will likewise be important. 
\section{Conclusions}
We observed the massive star-forming region W75N(B) with the VLBA to detect linearly and 
circularly polarized emission from 22~GHz \water ~masers associated with the two radio sources VLA\,1 and VLA\,2.
We observed that while the \water ~maser distribution and the magnetic field around VLA\,1 have not changed since 2005, the shell structure of the masers 
around VLA\,2 is still expanding and increasing its ellipticity. Furthermore, the magnetic field around VLA\,2 has changed its orientation according to 
the new direction of the major-axis of the shell-like structure and it is now aligned with the magnetic field in VLA\,1. We conclude that the 
\water ~masers around VLA\,2 are tracing the evolution  from a non-collimated to a collimated outflow.
\begin{acknowledgements} 
We wish to thank an anonymous referee for making useful suggestions that have improved the paper.
G.S. thanks Dr. D. Seifried for the useful discussion. J.M.T. acknowledges support from MICINN (Spain) grant AYA2011-30228-C03 
(co-funded with FEDER funds). The ICC (UB) is a CSIC-Associated Unit through the ICE (CSIC). Sc acknowledges support of DGAPA, UNAM, and CONACyT (M\'{e}xico).
 \end{acknowledgements}
\bibliographystyle{aa} 

\Online
\begin{appendix}

\section{Tables}
\label{appA}
\begin{figure*}[t!]
\centering
\includegraphics[width = 6.2 cm]{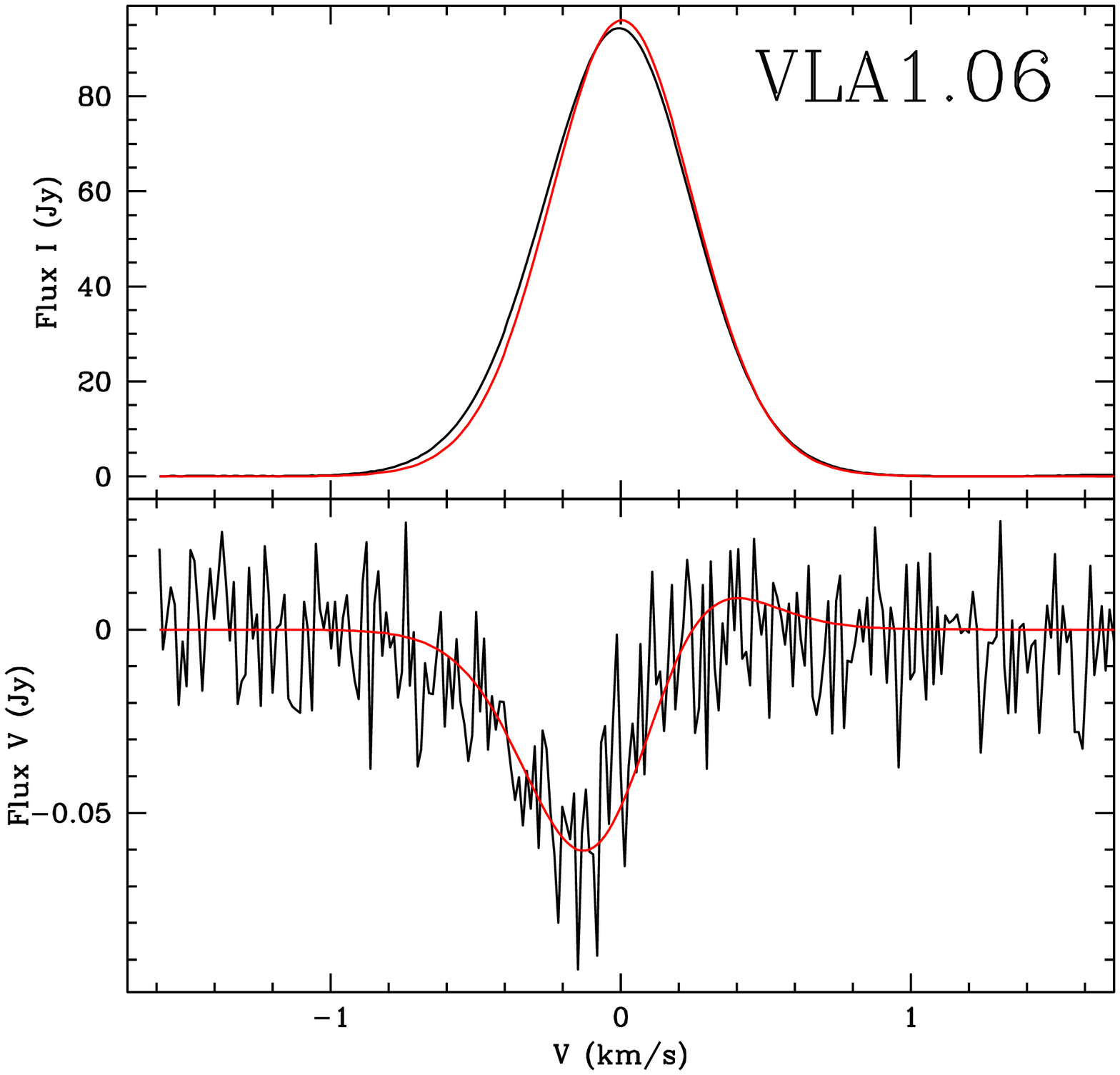}
\includegraphics[width = 6.2 cm]{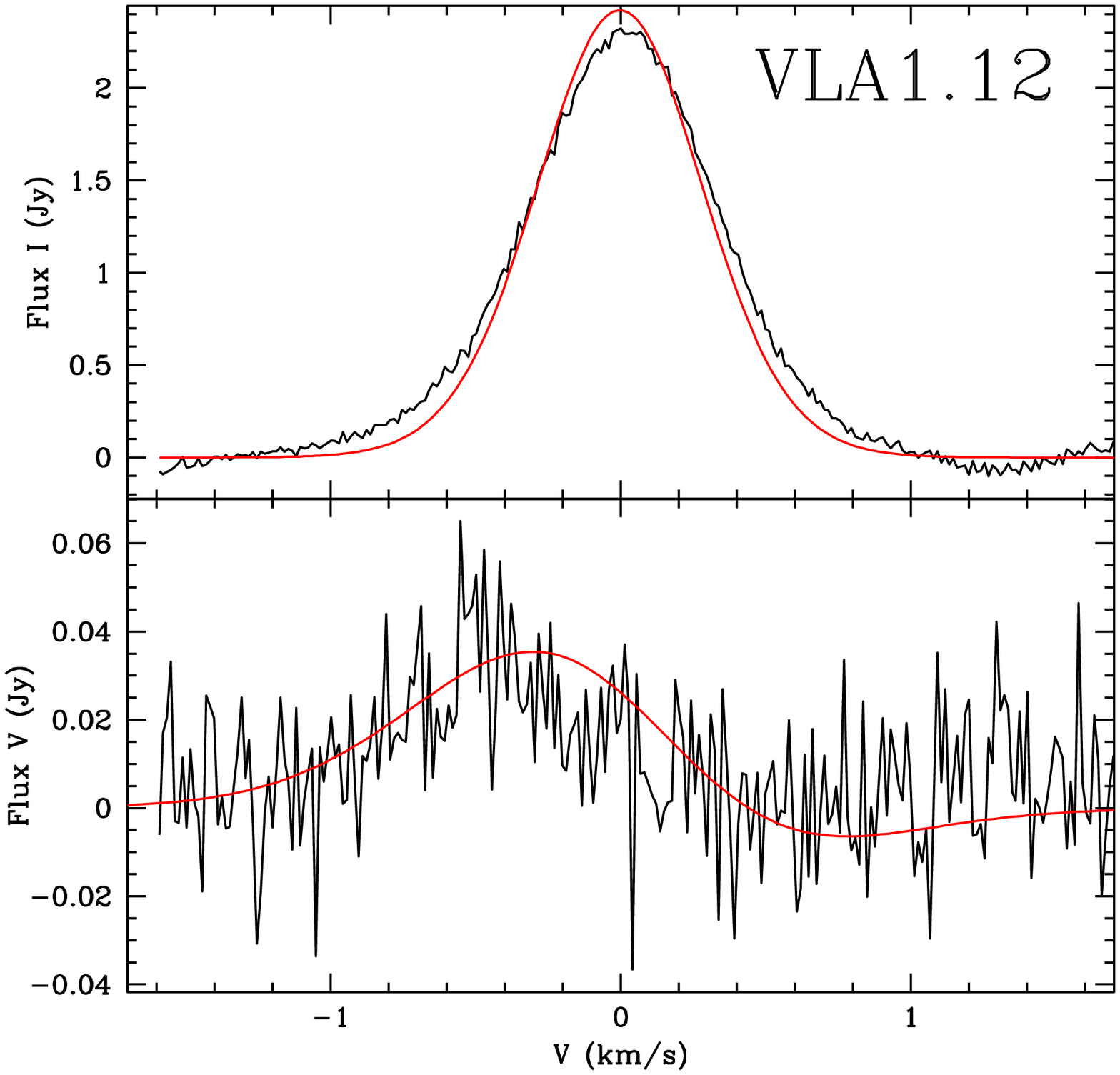}
\includegraphics[width = 6.2cm]{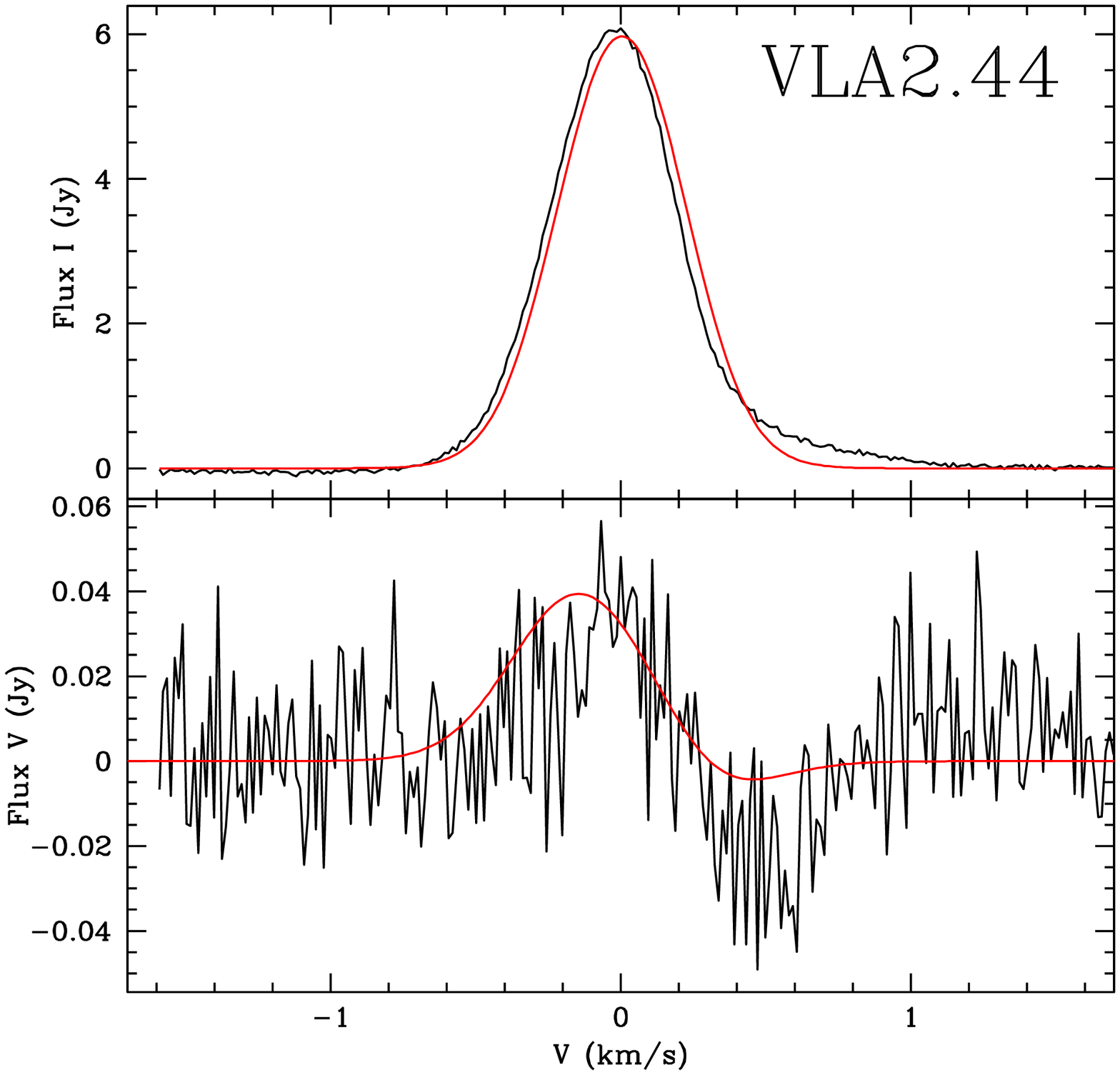}
\includegraphics[width = 6.2 cm]{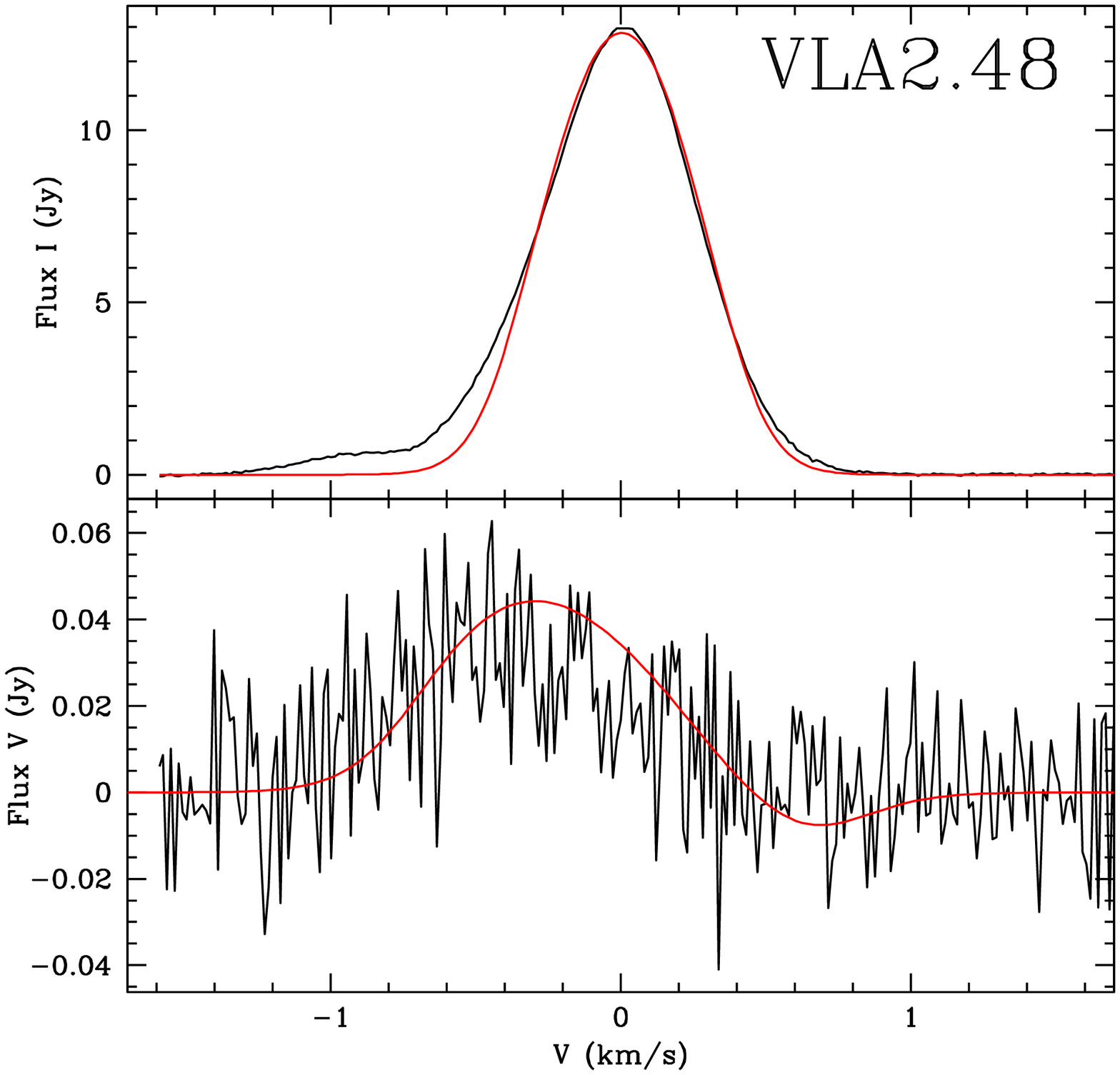}
\caption{Total intensity spectra (\textit{I}, \textit{upper panel}) and circular polarization intensity spectra (\textit{V}, 
\textit{lower panel}) for the \water ~masers VLA1.06, VLA1.12, VLA2.44, and VLA2.48 (see Tables~\ref{VLA1_wat} and \ref{VLA2_wat}).
The thick red line is the best-fit models of \textit{I} and \textit{V} emission obtained using the full radiative 
transfer method code for 22~GHz \water ~masers. The maser features were centered on zero velocity.}
\label{circpol}
\end{figure*}

In Tables~\ref{VLA1_wat} and \ref{VLA2_wat} we list all the \water ~maser features detected towards the two YSOs, VLA\,1 and VLA\,2, respectively. 
The tables are organized as follows. The name of the feature is reported in Col.~1. The positions, 
Cols.~2 and 3, refer to the brightest \water ~maser feature VLA1.06 that was used to self-calibrate the data. We estimated the 
absolute position of VLA1.06 to be  $\alpha_{2000}=20^{\rm{h}}38^{\rm{m}}36^{\rm{s}}\!.435$ and $\delta_{2000}=42^{\circ}37'34''\!\!.84$ (see Sect.~\ref{obsana}).
The peak flux density (I), the LSR velocity ($V_{\rm{lsr}}$), and the FWHM ($\Delta v\rm{_{L}}$) of the total intensity spectra of the 
maser features are reported in Cols.~4, 5, and 6, respectively; I, $V_{\rm{lsr}}$, and $\Delta v\rm{_{L}}$ are obtained using a Gaussian fit.
The mean linear polarization fraction ($P_{\rm{l}}$) and the mean linear polarization angles ($\chi$) are instead reported in Cols.~8 and 9, 
respectively. We determined $P_{\rm{l}}$ and $\chi$ of each \water ~maser feature by only considering the consecutive channels 
(more than two) across the total intensity spectrum for which the polarized intensity is $\geq5\sigma$.\\
\indent In Cols.~9 and 10 are reported
the values of the product of the brightness temperature $T_{\rm{b}}$ of the continuum radiation that is incident onto the masing 
region and the solid angle of the maser beam $\Delta \Omega$, which is known as the emerging brightness temperature \tbo, and the intrinsic 
thermal linewidth of the maser $\Delta V_{\rm{i}}$. Their values listed in the tables are the outputs of the FRTM code (Vlemmings et al. \cite{vle06}) 
that is based 
on the model for 22~GHz \water ~maser of Nedoluha \& Watson (\cite{ned92}), for which the shapes of the total intensity, linear polarization, and
circular polarization spectra depend on \tbo ~and \dvi ~(Nedoluha \& Watson \cite{ned91, ned92}). We model the observed linear polarized and total 
intensity maser spectra by gridding $\Delta V_{i}$ between 0.4~\kms and 4.0~\kms, in steps 
of 0.025~\kms, using a least-squares fitting routine ($\chi^{2}$-model) with $10^{6}$~K~sr~$<T_{\rm{b}}\Delta\Omega<10^{11}$~K~sr. We also set 
in our fit $(\Gamma+\Gamma_{\nu})=1 \rm{s^{-1}}$, where $\Gamma$ is the maser decay rate and $\Gamma_{\nu}$ is the 
cross-relaxation rate for the magnetic substated (see Vlemmings et al. \cite{vle06} and S11 for more details). \\
\indent From the maser theory we know that $P_{\rm{l}}$ 
of the \water ~maser emission depends on the degree of its saturation and the angle 
between the maser propagation direction and the magnetic field ($\theta$; e.g., Goldreich et al. \cite{gol73}). Because \tbo ~determines the 
relation between $P_{\rm{l}}$ and $\theta$, from the outputs of the FRTM code we are able to estimate the angles $\theta$ that are reported in
Col.~13. The errors of \tbo, \dvi, and $\theta$ are determined by analyzing the full probability distribution function. \\
\indent Finally, the best estimates of $T_{\rm{b}}\Delta\Omega$ and $\Delta V_{\rm{i}}$ are then included in the FRTM code to produce the 
\textit{I} and \textit{V} models that are used for fitting the total intensity and circular polarized spectra of the \water
~masers (see Fig.~\ref{circpol}). The magnetic field strength along the line of sight, which is reported in Col.~12, is finally evaluated by 
using the equation
\begin{equation}
B_{||}=B~cos \theta= \frac{P_{V}~\Delta v_{L}}{2 \cdot A_{F-F'}},
\label{magn}
\end{equation}
where $\Delta v_{L}$ is the FWHM of the total intensity spectrum, $P_{\rm{V}}=(V_{\rm{max}}-V_{\rm{max}})/I_{\rm{max}}$ is the circular 
polarization fraction (Col.~11), and the $A_{\rm{F-F'}}$ coefficient, which depends on $T_{\rm{b}}\Delta\Omega$, 
describes the relation between the circular polarization and the magnetic field strength for a transition between a high ($F$) and low ($F'$)
rotational energy level (Vlemmings et al. \cite{vle06}). \\
\indent In Table~\ref{wat0512} we compare the  parameters of the 22~GHz \water ~masers detected around VLA\,1 and VLA\,2 in epochs~2005.89 and 
2012.54. The first three rows are the observed parameters. In Rows~4 and 5 are reported the measured linear ($P_{\rm{l}}$) and circular polarization 
fraction ($P_{\rm{V}}$) in percentage. In the rest of the table we compare the intrinsic charateristics of the masers and the magnetic field properties that 
have all been estimated from the outputs of the FRTM code.
\begin {table*}[th!]
\caption []{All 22~GHz \water ~maser features detected around VLA\,1 (epoch~2012.54).} 
\begin{center}
\scriptsize
\begin{tabular}{ l c c c c c c c c c c c c}
\hline
\hline
\,\,\,\,\,(1)&(2)    & (3)      & (4)            & (5)       & (6)                 & (7)         & (8)       & (9)                     & (10)                        & (11)         & (12)                 &(13)                       \\
Maser     & RA       & Dec      & Peak flux      & $V_{\rm{lsr}}$ & $\Delta v\rm{_{L}}$ &$P_{\rm{l}}$ &  $\chi$   & $\Delta V_{\rm{i}}$ & $T_{\rm{b}}\Delta\Omega$\tablefootmark{a}& $P_{\rm{V}}$ & $B_{||}$  & $\theta$\\
          & offset   & offset   & Density(I)     &           &                     &             &	          &                         &                             &              &                      &       \\ 
          & (mas)    & (mas)    & (Jy/beam)      &  (km/s)   &      (km/s)         & (\%)        &   (\d)    & (km/s)                  & (log K sr)                  &   ($\%$)     &  (mG)               &  (\d)       \\ 
\hline
VLA1.01   & -101.297 & -188.122 & $1.64\pm0.16$  &  12.386   &      $0.47$         & $-$         & $-$       &  $-$                    & $-$                         & $-$	         & $-$                  & $-$    \\ 
VLA1.02   & -99.150  & -191.898 & $1.38\pm0.17$  &  12.305   &      $0.44$         & $-$         & $-$       &  $-$                    & $-$                         & $-$	         & $-$                  & $-$    \\ 
VLA1.03   & -62.311  & -218.605 & $2.85\pm0.12$  &  12.305   &      $0.44$         & $-$         & $-$       &  $-$                    & $-$                         & $-$	         & $-$                  & $-$    \\ 
VLA1.04   & -50.775  & -176.990 & $0.69\pm0.04$  &  11.861   &      $0.42$         & $-$         & $-$       &  $-$                    & $-$                         & $-$	         & $-$                  & $-$    \\ 
VLA1.05   & -11.662  & -39.768  & $9.80\pm0.04$  &  11.928   &      $0.53$         & $0.9\pm0.1$ & $-30\pm5$ &  $-$                    & $-$                         & $-$	         & $-$                  & $-$    \\ 
VLA1.06   & 0        & 0        & $94.32\pm0.11$ &  10.593   &      $0.60$         & $1.5\pm0.4$ & $-26\pm6$ &  $<2.0$                 & $<6.00$                     & $0.07$	         & $+18\pm6$            & $90^{+7}_{-7}$    \\ 
VLA1.07   & 18.020   & -66.166  & $2.19\pm0.06$  &   9.501   &      $0.51$         & $2.8\pm0.3$ & $-25\pm8$ &  $<1.2$                 & $<6.00$                     & $-$	         & $-$                  & $90^{+29}_{-29}$    \\ 
VLA1.08   & 19.114   & -66.052  & $1.27\pm0.03$  &   9.811   &      $0.69$         & $-$         & $-$       &  $-$                    & $-$                         & $-$	         & $-$                  & $-$    \\ 
VLA1.09   & 19.493   & -64.823  & $0.83\pm0.03$  &   9.811   &      $0.52$         & $-$         & $-$       &  $-$                    & $-$                         & $-$	         & $-$                  & $-$    \\ 
VLA1.10   & 23.998   & 18.848   & $4.82\pm0.06$  &  10.957   &      $0.77$         & $2.8\pm0.2$ & $-52\pm4$ &  $<2.4$                 & $<6.00$                     & $-$	         & $-$                  & $90^{+8}_{-8}$    \\ 
VLA1.11   & 24.335   & 17.906   & $1.68\pm0.05$  &  11.820   &      $0.65$         & $-$         & $-$       &  $-$                    & $-$                         & $-$	         & $-$                  & $-$    \\ 
VLA1.12   & 30.313   & 18.917   & $2.33\pm0.06$  &  11.159   &      $0.74$         & $4.5\pm1.4$ & $-70\pm7$ &  $-$                    & $-$                         & $1.8\tablefootmark{b}$ & $-544\pm272\tablefootmark{b}$         & $-$    \\ 
VLA1.13   & 31.787   & -61.127  & $2.50\pm0.06$  &  12.872   &      $1.06$         & $-$         & $-$       &  $-$                    & $-$                         & $-$	         & $-$                  & $-$    \\ 
VLA1.14   & 32.040   & -61.161  & $1.01\pm0.04$  &  13.977   &      $1.14$         & $-$         & $-$       &  $-$                    & $-$                         & $-$	         & $-$                  & $-$    \\ 
VLA1.15   & 32.166   & 19.795   & $1.28\pm0.06$  &  10.930   &      $0.70$         & $-$         & $-$       &  $-$                    & $-$                         & $-$	         & $-$                  & $-$    \\ 
VLA1.16   & 36.587   & 11.280   & $0.90\pm0.07$  &  14.611   &      $1.05$         & $-$         & $-$       &  $-$                    & $-$                         & $-$	         & $-$                  & $-$    \\ 
VLA1.17   & 36.671   & -133.79  & $1.10\pm0.03$  &   9.784   &      $0.34$         & $-$         & $-$       &  $-$                    & $-$                         & $-$	         & $-$                  & $-$    \\ 
VLA1.18   & 42.776   & 22.419   & $1.07\pm0.06$  &  11.267   &      $0.64$         & $-$         & $-$       &  $-$                    & $-$                         & $-$	         & $-$                  & $-$    \\ 
VLA1.19   & 43.449   & 22.812   & $1.08\pm0.05$  &  11.362   &      $0.93$         & $-$         & $-$       &  $-$                    & $-$                         & $-$	         & $-$                  & $-$    \\ 
VLA1.20   & 57.385   & 35.961   & $3.95\pm0.03$  &   8.652   &      $0.78$         & $-$         & $-$       &  $-$                    & $-$                         & $-$	         & $-$                  & $-$    \\ 
VLA1.21   & 62.311   & 36.697   & $0.81\pm0.05$  &  11.389   &      $1.01$         & $-$         & $-$       &  $-$                    & $-$                         & $-$	         & $-$                  & $-$    \\ 
VLA1.22   & 63.406   & 40.180   & $4.27\pm0.03$  &   9.933   &      $0.63$         & $-$         & $-$       &  $-$                    & $-$                         & $-$	         & $-$                  & $-$    \\ 
VLA1.23   & 79.615   & 48.130   & $16.48\pm0.06$ &  12.845   &      $1.77$         & $0.6\pm0.1$ & $-38\pm5$ &  $-$                    & $-$                         & $-$	         & $-$                  & $-$    \\ 
VLA1.24   & 80.836   & 49.847   & $10.09\pm0.07$ &  12.764   &      $1.16$         & $0.8\pm0.1$ & $-40\pm6$ &  $<3.6$                 & $<6.00$                     & $-$	         & $-$                  & $90^{+8}_{-8}$    \\ 
VLA1.25   & 81.131   & 45.647   & $1.52\pm0.05$  &  11.267   &      $0.78$         & $-$         & $-$       &  $-$                    & $-$                         & $-$	         & $-$                  & $-$    \\ 
VLA1.26   & 81.299   & 49.385   & $7.80\pm0.04$  &  13.640   &      $2.10$         & $-$         & $-$       &  $-$                    & $-$                         & $-$	         & $-$                  & $-$    \\ 
VLA1.27   & 82.141   & 49.545   & $1.93\pm0.12$  &  15.703   &      $0.87$         & $-$         & $-$       &  $-$                    & $-$                         & $-$	         & $-$                  & $-$    \\ 
VLA1.28   & 82.520   & 46.993   & $1.22\pm0.05$  &  11.348   &      $1.60$         & $-$         & $-$       &  $-$                    & $-$                         & $-$	         & $-$                  & $-$    \\ 
VLA1.29   & 82.562   & 46.635   & $1.65\pm0.22$  &  15.298   &      $1.75$         & $-$         & $-$       &  $-$                    & $-$                         & $-$	         & $-$                  & $-$    \\ 
VLA1.30   & 83.194   & 47.424   & $0.45\pm0.04$  &  10.148   &      $1.47$         & $-$         & $-$       &  $-$                    & $-$                         & $-$	         & $-$                  & $-$    \\ 
VLA1.31   & 88.246   & 19.569   & $4.00\pm0.09$  &   9.407   &      $1.53$         & $-$         & $-$       &  $-$                    & $-$                         & $-$	         & $-$                  & $-$    \\ 
VLA1.32   & 89.383   & 18.864   & $4.61\pm0.05$  &  11.038   &      $2.66$         & $-$         & $-$       &  $-$                    & $-$                         & $-$	         & $-$                  & $-$    \\ 
VLA1.33   & 89.719   & 19.123   & $3.97\pm0.13$  &  12.494   &      $1.64$         & $-$         & $-$       &  $-$                    & $-$                         & $-$	         & $-$                  & $-$    \\ 
VLA1.34   & 89.762   & 21.153   & $1.72\pm0.08$  &  12.090   &      $1.99$         & $-$         & $-$       &  $-$                    & $-$                         & $-$	         & $-$                  & $-$    \\ 
VLA1.35   & 89.804   & 19.898   & $4.58\pm0.04$  &  13.317   &      $0.78$         & $-$         & $-$       &  $-$                    & $-$                         & $-$	         & $-$                  & $-$    \\ 
VLA1.36   & 90.140   & 19.264   & $2.24\pm0.09$  &  10.391   &      $3.80$         & $-$         & $-$       &  $-$                    & $-$                         & $-$	         & $-$                  & $-$    \\ 
VLA1.37   & 91.109   & 16.884   & $1.67\pm0.05$  &  14.018   &      $0.92$         & $-$         & $-$       &  $-$                    & $-$                         & $-$	         & $-$                  & $-$    \\ 
VLA1.38   & 94.098   & 8.427    & $0.61\pm0.05$  &  11.537   &      $1.21$         & $-$         & $-$       &  $-$                    & $-$                         & $-$	         & $-$                  & $-$    \\ 
\hline
\end{tabular} 
\end{center}
\tablefoot{
\tablefoottext{a}{The output \tbo ~must be adjusted according to 
the real value of $\Gamma+\Gamma_{\nu}$, which depends on the gas temperature ($T$). Using $\Delta V_{\rm{i}}\approx 0.5~(T/100)^{1/2}$ 
(Vlemmings et al. \cite{vle06}) we estimated that $T_{\rm{VLA1}}<2300$~K for which \tbo ~has to be adjusted by adding at most $+1.11$~log~K~sr 
(Anderson \& Watson \cite{and93}).}
\tablefoottext{b}{In the fitting model we include the values \tbo$=1\times10^6$~K~sr and \dvi$=2.4$~\kms ~that are the estimated upper limits of VLA1.10 (see Fig.~\ref{circpol}).}
}
\label{VLA1_wat}
\end{table*}
\begin {table*}[]
\caption []{All 22~GHz \water ~maser features detected around VLA\,2 (epoch~2012.54).} 
\begin{center}
\scriptsize
\begin{tabular}{ l c c c c c c c c c c c c}
\hline
\hline
\,\,\,\,\,(1)&(2)    & (3)      & (4)            & (5)       & (6)              & (7)         & (8)       & (9)                     & (10)                        & (11)         & (12)                 &(13)                       \\
Maser     & RA\tablefootmark{a}& Dec\tablefootmark{a}& Peak flux & $V_{\rm{lsr}}$& $\Delta v\rm{_{L}}$ &$P_{\rm{l}}$&$\chi$ & $\Delta V_{\rm{i}}$  & $T_{\rm{b}}\Delta\Omega$\tablefootmark{a}& $P_{\rm{V}}$ & $B_{||}$  & $\theta$\\
          & offset   & offset   & Density(I)     &           &                  &             &	                            &                         &                             &              &                      &       \\ 
          & (mas)    & (mas)    & (Jy/beam)      &  (km/s)   &      (km/s)      & (\%)        &   (\d)                      & (km/s)                  & (log K sr)                  &   ($\%$)     &  (mG)               &  (\d)       \\ 
\hline
VLA2.01   & 473.816  & -727.013 & $1.24\pm0.05$  &  16.309   &      $0.54$      & $-$         & $-$       &  $-$                    & $-$                         & $-$	         & $-$                  & $-$    \\ 
VLA2.02   & 474.784  & -730.545 & $0.48\pm0.03$  &  16.512   &      $0.73$      & $-$         & $-$       &  $-$                    & $-$                         & $-$	         & $-$                  & $-$    \\ 
VLA2.03   & 481.352  & -711.754 & $1.20\pm0.03$  &  16.498   &      $0.79$      & $-$         & $-$       &  $-$                    & $-$                         & $-$	         & $-$                  & $-$    \\ 
VLA2.04   & 485.984  & -710.171 & $0.22\pm0.02$  &  20.044   &      $0.64$      & $-$         & $-$       &  $-$                    & $-$                         & $-$	         & $-$                  & $-$    \\ 
VLA2.05   & 487.962  & -811.104 & $6.62\pm0.04$  &  14.233   &      $0.43$      & $1.6\pm0.1$ & $-12\pm4$ &  $-$                    & $-$                         & $-$	         & $-$                  & $-$    \\ 
VLA2.06   & 491.709  & -813.095 & $0.57\pm0.04$  &  13.384   &      $0.53$      & $-$         & $-$       &  $-$                    & $-$                         & $-$	         & $-$                  & $-$    \\ 
VLA2.07   & 498.530  & -761.936 & $0.35\pm0.03$  &  11.982   &      $0.73$      & $-$         & $-$       &  $-$                    & $-$                         & $-$	         & $-$                  & $-$    \\ 
VLA2.08   & 519.370  & -737.568 & $0.97\pm0.02$  &   6.670   &      $0.72$      & $-$         & $-$       &  $-$                    & $-$                         & $-$	         & $-$                  & $-$    \\ 
VLA2.09   & 519.539  & -707.607 & $0.69\pm0.02$  &  18.898   &      $0.79$      & $-$         & $-$       &  $-$                    & $-$                         & $-$	         & $-$                  & $-$    \\ 
VLA2.10   & 519.749  & -708.904 & $3.24\pm0.02$  &  18.588   &      $0.56$      & $-$         & $-$       &  $-$                    & $-$                         & $-$	         & $-$                  & $-$    \\ 
VLA2.11   & 528.507  & -660.904 & $0.21\pm0.02$  &   2.369   &      $0.55$      & $-$         & $-$       &  $-$                    & $-$                         & $-$	         & $-$                  & $-$    \\ 
VLA2.12   & 529.643  & -659.538 & $0.22\pm0.02$  &   2.504   &      $0.57$      & $-$         & $-$       &  $-$                    & $-$                         & $-$	         & $-$                  & $-$    \\ 
VLA2.13   & 560.883  & -794.949 & $6.01\pm0.03$  &  11.645   &      $0.53$      & $-$         & $-$       &  $-$                    & $-$                         & $-$	         & $-$                  & $-$    \\ 
VLA2.14   & 572.629  & -591.267 & $0.84\pm0.02$  &   3.798   &      $0.66$      & $-$         & $-$       &  $-$                    & $-$                         & $-$	         & $-$                  & $-$    \\ 
VLA2.15   & 574.061  & -591.526 & $1.03\pm0.02$  &   4.648   &      $0.57$      & $-$         & $-$       &  $-$                    & $-$                         & $-$	         & $-$                  & $-$    \\ 
VLA2.16   & 574.945  & -801.651 & $3.76\pm0.18$  &  15.110   &      $0.64$      & $-$         & $-$       &  $-$                    & $-$                         & $-$	         & $-$                  & $-$    \\ 
VLA2.17   & 575.871  & -800.690 & $3.02\pm0.12$  &  15.420   &      $1.05$      & $-$         & $-$       &  $-$                    & $-$                         & $-$	         & $-$                  & $-$    \\ 
VLA2.18   & 576.250  & -800.716 & $2.16\pm0.15$  &  14.988   &      $1.24$      & $-$         & $-$       &  $-$                    & $-$                         & $-$	         & $-$                  & $-$    \\ 
VLA2.19   & 578.397  & -799.526 & $8.05\pm0.05$  &  14.409   &      $0.62$      & $1.1\pm0.1$ & $-46\pm5$ &  $-$                    & $-$                         & $-$	         & $-$                  & $-$    \\ 
VLA2.20   & 578.945  & -594.635 & $0.32\pm0.03$  &   6.158   &      $0.62$      & $-$         & $-$       &  $-$                    & $-$                         & $-$	         & $-$                  & $-$    \\ 
VLA2.21   & 579.829  & -796.158 & $0.27\pm0.03$  &  17.334   &      $1.10$      & $-$         & $-$       &  $-$                    & $-$                         & $-$	         & $-$                  & $-$    \\ 
VLA2.22   & 580.039  & -594.902 & $0.29\pm0.03$  &   6.104   &      $0.60$      & $-$         & $-$       &  $-$                    & $-$                         & $-$	         & $-$                  & $-$    \\ 
VLA2.23   & 582.313  & -610.703 & $0.60\pm0.02$  &   7.776   &      $0.54$      & $-$         & $-$       &  $-$                    & $-$                         & $-$	         & $-$                  & $-$    \\ 
VLA2.24   & 583.197  & -602.474 & $1.14\pm0.02$  &   4.648   &      $0.57$      & $-$         & $-$       &  $-$                    & $-$                         & $-$	         & $-$                  & $-$    \\ 
VLA2.25   & 585.555  & -793.598 & $2.79\pm0.13$  &  14.867   &      $1.10$      & $-$         & $-$       &  $-$                    & $-$                         & $-$	         & $-$                  & $-$    \\ 
VLA2.26   & 586.481  & -793.186 & $1.75\pm0.03$  &  13.613   &      $1.22$      & $-$         & $-$       &  $-$                    & $-$                         & $-$	         & $-$                  & $-$    \\ 
VLA2.27   & 588.755  & -793.449 & $6.04\pm0.15$  &  15.298   &      $0.84$      & $-$         & $-$       &  $-$                    & $-$                         & $-$	         & $-$                  & $-$    \\ 
VLA2.28   & 594.817  & -599.632 & $1.52\pm0.02$  &   7.371   &      $0.58$      & $-$         & $-$       &  $-$                    & $-$                         & $-$	         & $-$                  & $-$    \\ 
VLA2.29   & 602.522  & -739.132 & $0.26\pm0.02$  &  22.039   &      $0.89$      & $-$         & $-$       &  $-$                    & $-$                         & $-$	         & $-$                  & $-$    \\ 
VLA2.30   & 604.205  & -740.036 & $2.84\pm0.02$  &  20.758   &      $0.71$      & $-$         & $-$       &  $-$                    & $-$                         & $-$	         & $-$                  & $-$    \\ 
VLA2.31   & 606.016  & -740.799 & $0.44\pm0.02$  &  18.696   &      $1.20$      & $-$         & $-$       &  $-$                    & $-$                         & $-$	         & $-$                  & $-$    \\ 
VLA2.32   & 607.153  & -741.844 & $0.71\pm0.02$  &  19.316   &      $0.81$      & $-$         & $-$       &  $-$                    & $-$                         & $-$	         & $-$                  & $-$    \\ 
VLA2.33   & 608.079  & -742.798 & $0.73\pm0.02$  &  19.248   &      $1.35$      & $-$         & $-$       &  $-$                    & $-$                         & $-$	         & $-$                  & $-$    \\ 
VLA2.34   & 608.416  & -743.553 & $0.91\pm0.02$  &  19.976   &      $0.92$      & $-$         & $-$       &  $-$                    & $-$                         & $-$	         & $-$                  & $-$    \\ 
VLA2.35   & 610.732  & -744.667 & $3.19\pm0.02$  &  18.494   &      $0.75$      & $-$         & $-$       &  $-$                    & $-$                         & $-$	         & $-$                  & $-$    \\ 
VLA2.36   & 613.847  & -744.747 & $0.48\pm0.05$  &  10.984   &      $0.87$      & $-$         & $-$       &  $-$                    & $-$                         & $-$	         & $-$                  & $-$    \\ 
VLA2.37   & 614.605  & -770.691 & $0.64\pm0.04$  &  13.249   &      $0.72$      & $-$         & $-$       &  $-$                    & $-$                         & $-$	         & $-$                  & $-$    \\ 
VLA2.38   & 614.647  & -772.896 & $0.35\pm0.03$  &  13.573   &      $0.79$      & $-$         & $-$       &  $-$                    & $-$                         & $-$	         & $-$                  & $-$    \\ 
VLA2.39   & 614.774  & -768.429 & $0.34\pm0.04$  &  12.535   &      $0.90$      & $-$         & $-$       &  $-$                    & $-$                         & $-$	         & $-$                  & $-$    \\ 
VLA2.40   & 652.539  & -747.784 & $0.24\pm0.02$  &  -2.191   &      $0.90$      & $-$         & $-$       &  $-$                    & $-$                         & $-$	         & $-$                  & $-$    \\ 
VLA2.41   & 653.465  & -572.544 & $0.29\pm0.02$  &   7.506   &      $1.00$      & $-$         & $-$       &  $-$                    & $-$                         & $-$	         & $-$                  & $-$    \\ 
VLA2.42   & 659.149  & -586.201 & $0.44\pm0.02$  & -10.155   &      $0.75$      & $-$         & $-$       &  $-$                    & $-$                         & $-$	         & $-$                  & $-$    \\ 
VLA2.43   & 685.042  & -666.283 & $1.08\pm0.02$  &   2.329   &      $1.09$      & $-$         & $-$       &  $-$                    & $-$                         & $-$	         & $-$                  & $-$    \\ 
VLA2.44   & 685.084  & -601.471 & $6.08\pm0.14$  &  14.934   &      $0.52$      & $-$         & $-$       &  $-$                    & $-$                         & $0.7$\tablefootmark{b} & $-152\pm73$\tablefootmark{b}          & $-$    \\ 
VLA2.45   & 686.305  & -605.049 & $8.87\pm0.14$  &  15.325   &      $0.49$      & $1.2\pm0.2$ & $-4\pm7$  &  $-$                    & $-$                         & $-$	         & $-$                  & $-$    \\ 
VLA2.46   & 687.610  & -655.483 & $4.26\pm0.02$  &  -6.636   &      $0.64$      & $-$         & $-$       &  $-$                    & $-$                         & $-$	         & $-$                  & $-$    \\ 
VLA2.47   & 687.947  & -606.522 & $13.84\pm0.15$ &  15.285   &      $0.44$      & $-$         & $-$       &  $-$                    & $-$                         & $-$	         & $-$                  & $-$    \\ 
VLA2.48   & 688.157  & -609.310 & $12.97\pm0.07$ &  16.094   &      $0.60$      & $-$         & $-$       &  $-$                    & $-$                         & $0.4$\tablefootmark{c} & $-103\pm43$\tablefootmark{c}          & $-$    \\ 
VLA2.49   & 688.199  & -672.810 & $0.27\pm0.02$  &   1.776   &      $0.63$      & $-$         & $-$       &  $-$                    & $-$                         & $-$	         & $-$                  & $-$    \\ 
VLA2.50   & 688.831  & -647.507 & $0.38\pm0.02$  &  -3.158   &      $0.56$      & $-$         & $-$       &  $-$                    & $-$                         & $-$	         & $-$                  & $-$    \\ 
VLA2.51   & 692.115  & -619.831 & $4.77\pm0.04$  &  12.777   &      $0.52$      & $-$         & $-$       &  $-$                    & $-$                         & $-$	         & $-$                  & $-$    \\ 
VLA2.52   & 699.188  & -636.635 & $14.07\pm0.17$ &  15.002   &      $0.59$      & $0.7\pm0.1$ & $-45\pm6$ &  $-$                    & $-$                         & $-$	         & $-$                  & $-$    \\ 
VLA2.53   & 699.609  & -659.790 & $9.75\pm0.03$  &  17.213   &      $0.58$      & $-$         & $-$       &  $-$                    & $-$                         & $-$	         & $-$                  & $-$    \\ 
VLA2.54   & 702.514  & -655.918 & $0.65\pm0.04$  &  16.256   &      $0.55$      & $-$         & $-$       &  $-$                    & $-$                         & $-$	         & $-$                  & $-$    \\ 
VLA2.55   & 702.809  & -657.299 & $1.11\pm0.05$  &  16.269   &      $0.46$      & $-$         & $-$       &  $-$                    & $-$                         & $-$	         & $-$                  & $-$    \\ 
VLA2.56   & 703.398  & -636.490 & $1.16\pm0.06$  &  14.584   &      $0.42$      & $-$         & $-$       &  $-$                    & $-$                         & $-$	         & $-$                  & $-$    \\ 
VLA2.57   & 703.483  & -658.810 & $9.30\pm0.07$  &  15.851   &      $0.46$      & $-$         & $-$       &  $-$                    & $-$                         & $-$	         & $-$                  & $-$    \\ 
VLA2.58   & 710.514  & -668.293 & $0.26\pm0.02$  &  -6.219   &      $0.65$      & $-$         & $-$       &  $-$                    & $-$                         & $-$	         & $-$                  & $-$    \\ 
VLA2.59   & 710.556  & -926.731 & $4.29\pm0.03$  &   8.179   &      $0.72$      & $-$         & $-$       &  $-$                    & $-$                         & $-$	         & $-$                  & $-$    \\ 
VLA2.60   & 711.987  & -661.125 & $3.56\pm0.02$  &   6.764   &      $1.09$      & $-$         & $-$       &  $-$                    & $-$                         & $-$	         & $-$                  & $-$    \\ 
VLA2.61   & 712.282  & -660.404 & $1.14\pm0.02$  &   8.180   &      $0.88$      & $-$         & $-$       &  $-$                    & $-$                         & $-$	         & $-$                  & $-$    \\ 
VLA2.62   & 712.282  & -658.833 & $0.88\pm0.06$  &  10.243   &      $0.62$      & $-$         & $-$       &  $-$                    & $-$                         & $-$	         & $-$                  & $-$    \\ 
VLA2.63   & 712.450  & -661.923 & $10.81\pm0.03$ &   5.929   &      $0.83$      & $-$         & $-$       &  $-$                    & $-$                         & $-$	         & $-$                  & $-$    \\ 
VLA2.64   & 714.597  & -925.274 & $26.25\pm0.04$ &   9.232   &      $0.53$      & $0.7\pm0.1$ & $-48\pm3$ &  $1.98^{+0.05}_{-0.17}$ & $8.8^{+0.3}_{-0.1}$         & $-$	         & $-$                  & $84^{+6}_{-10} $    \\ 
VLA2.65   & 716.071  & -671.002 & $0.19\pm0.02$  &  -7.449   &      $0.89$      & $-$         & $-$       &  $-$                    & $-$                         & $-$	         & $-$                  & $-$    \\ 
VLA2.66   & 719.018  & -670.666 & $0.26\pm0.02$  &   5.308   &      $0.79$      & $-$         & $-$       &  $-$                    & $-$                         & $-$	         & $-$                  & $-$    \\ 
VLA2.67   & 740.785  & -707.310 & $0.90\pm0.02$  & -11.193   &      $0.59$      & $-$         & $-$       &  $-$                    & $-$                         & $-$	         & $-$                  & $-$    \\ 
VLA2.68   & 742.258  & -708.374 & $0.24\pm0.02$  & -11.166   &      $0.51$      & $-$         & $-$       &  $-$                    & $-$                         & $-$	         & $-$                  & $-$    \\
\hline
\end{tabular} 
\end{center}
\tablefoot{
\tablefoottext{a}{The output \tbo ~must be adjusted according to 
the real value of $\Gamma+\Gamma_{\nu}$, which depends on the gas temperature ($T$). Using $\Delta V_{\rm{i}}\approx 0.5~(T/100)^{1/2}$ 
(Vlemmings et al. \cite{vle06}) we estimated that $T_{\rm{VLA\,2}}=1600$~K ($\Gamma+\Gamma_{\nu}=9~\rm{s}^{-1}$) for which \tbo ~has to be adjusted 
by adding $+0.95$~log~K~sr (Anderson \& Watson \cite{and93}).}
\tablefoottext{b}{In the fitting model we include the values \tbo$=5\times10^9$~K~sr and \dvi$=2.0$~\kms.}
\tablefoottext{c}{In the fitting model we include the values \tbo$=10^9$~K~sr and \dvi$=2.0$~\kms.}
}
\label{VLA2_wat}
\end{table*}

\begin {table*}[t!]
\caption []{Comparison of 22~GHz \water ~maser parameters between epochs~2005.89 (S11) and 2012.54 (this work).} 
\begin{center}
\scriptsize
\begin{tabular}{ l c c c c }
\hline
\hline
                                         & \multicolumn{2}{c}{epoch 2005.89\tablefootmark{a}} & \multicolumn{2}{c}{epoch 2012.54}     \\ 
                                         &  VLA\,1               &     VLA\,2            &        VLA\,1         &     VLA\,2        \\
\hline
Number of maser features                 &    36               &     88              &        38           &     68          \\
    $V_{\rm{lsr}}$ range (\kms)          & $[+8.1; +24.7]$     & $[-7.7; +14.8]$     & $[+9.4; +15.7]$     & $[-11.2; +22.0]$\\
$I$ range (\jyb)                         & $[0.13; 158.28]$    & $[0.09; 198.45]$    & $[0.45; 94.32]$     & $[0.19; 26.25]$ \\
\hline
                                         & \multicolumn{4}{c}{Polarization\tablefootmark{b}} \\
\hline
 $P_{\rm{l}}$ range (\%)                 & $[0.6; 25.7]$       & $[0.2; 6.1]$        & $[0.6; 4.5]$        & $[0.7; 1.6]$    \\
 $P_{\rm{V}}$ range (\%)                 & $[0.3; 4.2]$        & $[0.1; 3.0 ]$       & $[0.1; 1.8]$        & $[0.4; 0.7]$    \\
  \hline
                                         & \multicolumn{4}{c}{Intrinsic characteristics} \\
\hline
 $\Delta V_{\rm{i}}$ range (\kms)        & $[0.6; 2.0]$        & $[0.7; 3.4]$        & $[1.2; 3.6]$        & 2.0             \\
$T_{\rm{b}}\Delta\Omega$ range (log K sr)& $[8.9; 10.7]$       & $[8.6; 10.7]$       & $[6.0; 6.0]$        & 8.8             \\
 $\langle\Delta V_{\rm{i}}\rangle$\tablefootmark{b} (\kms)& $1.0^{+0.8}_{-0.2}$& $2.5^{+0.6}_{-0.4}$ & $<2.4$ & $2.0^{+0.1}_{-0.2}$ \\
$\langle T_{\rm{b}}\Delta\Omega\rangle$\tablefootmark{b} (log K sr)&$10.6^{+1.1}_{-0.5}$&$9.4^{+0.6}_{-0.9}$  &$<6.0$& $8.8^{+0.3}_{-0.1}$ \\
$T$\tablefootmark{d} (K)                 & $400$               & $2500$              & $<2300$              & $1600$             \\
$\Gamma+\Gamma_{\nu}$\tablefootmark{d}   & $3$                 & $14$                & $<13$                 & $9$             \\
\hline
                                         & \multicolumn{4}{c}{Magnetic field} \\
\hline
 $\chi$ range (\d)                       &  $[-90; +58]$       & $[-90; +83]$        & $[-70; -25]$     & $[-48; -4]$        \\
 $\theta$ range	(\d)			 &  $[+66; +90]$       & $[+61; +90]$        & $[+90; +90]$     & $+84$              \\
 $\Phi_{\rm{B}}$ range	(\d)		 &  $[-32; 0]$         & $[-7; 0]$           & $[+20; +65]$     & $[+42; +86]$       \\
 $|B_{||}|$ range (mG)                   &  $[54; 809]$        & $[38; 957]$         & $[18; 544]$      & $[103; 152]$       \\
 $\langle \chi \rangle$\tablefootmark{e} (\d) &  $-67\pm40$         & $-72\pm32$          & $-41\pm15$       & $-33\pm21$         \\
 $\langle \theta \rangle$\tablefootmark{e} (\d) &  $+83^{+7}_{-15}$   & $+85^{+6}_{-36}$    & $+90^{+10}_{-10}$& $+84^{+6}_{-10} $  \\
 $\langle\Phi_{\rm{B}}\rangle$ (\d)      &  $+23\pm40$         & $+18\pm32$          & $+49\pm15$       & $+57\pm21$         \\
 $\langle |B_{||}| \rangle$\tablefootmark{f} (mG) &  $81\pm62$ & $145\pm110$         & $>18$       & $116\pm35$         \\
 $\langle |B| \rangle$\tablefootmark{f,g} (mG) &  $665\pm509$  & $1663\pm1262$       & $>104$\tablefootmark{h} & $1110\pm335$       \\
 Arithmetic mean of $|B_{||}|$ (mG)      &  $264$              & $345$               & $281$            & $128$              \\
\hline
\end{tabular}
\end{center}
\tablefoot{
\tablefoottext{a}{Parameters from S11.}
\tablefoottext{b}{Averaged values determined by analyzing the total full probability distribution function.}
\tablefoottext{c}{$T\approx100 \cdot (\Delta V_{\rm{i}}/0.5)^2$ is the gas temperature of the region where the \water ~masers arise.}
\tablefoottext{d}{Cross-relaxation rate. The values of \tbo ~have to be adjusted by adding +0.48 ($\Gamma+\Gamma_{\nu}=3$), +0.95 ($\Gamma+\Gamma_{\nu}=9$),
+1.11 ($\Gamma+\Gamma_{\nu}=13$), +1.15 ($\Gamma+\Gamma_{\nu}=14$) as described in Anderson \& Watson (\cite{and93}).}
\tablefoottext{e}{Error-weighted values, where the weights are $w_{\rm{i}}=1/e_{\rm{i}}$ and $e_{\rm{i}}$ is the error
of the i\textit{th} measurements.}
\tablefoottext{f}{Error-weighted values, where the weights are $w_{\rm{i}}=1/e_{\rm{i}}^{2}$.}
\tablefoottext{g}{$|B|=\langle |B_{||}| \rangle/\rm{cos}\langle \theta \rangle$ if $\theta\neq\pm90$\d.}
\tablefoottext{h}{We report the lower limit estimated by considering $\theta=80$\d.}
}
\label{wat0512}
\end{table*}

\end{appendix}

\end{document}